\documentclass[%
 reprint,
 amsmath,amssymb,
 aps,
pre,
]{revtex4-2}

\usepackage{amsmath}
\usepackage{dcolumn}
\usepackage{bm}
\usepackage{braket}
\usepackage{mathtools}
\usepackage{tensor}
\usepackage{graphicx} 
\usepackage{bbm}
\usepackage{mleftright}
\allowdisplaybreaks
\usepackage{xcolor}

\newcommand{\nx}{N_X}
\newcommand{\ny}{N_Y}
\newcommand{\qx}{q_X}
\newcommand{\qy}{q_Y}
\newcommand{\px}{p_X}
\newcommand{\py}{p_Y}

\newcommand{\xmat}{\mathbf{X}}

\newcommand{\rxmat}{\mathbf{R}_X}
\newcommand{\vx}{\hat{v}_x}
\newcommand{\ymat}{\mathbf{Y}}

\newcommand{\rymat}{\mathbf{R}_Y}
\newcommand{\vy}{\hat{v}_y}

\newcommand{\latentmat}{\mathbf{u}}
\newcommand{\vz}{\hat{v}_z}
\newcommand{\vv}{\hat{v}}
\newcommand{\vmax}{\hat{v}_{\rm max}}
\newcommand{\vzjoint}{\hat{v}_{z, \mathrm{joint}}}
\newcommand{\vxjoint}{\hat{v}_{x, \mathrm{joint}}}
\newcommand{\vyjoint}{\hat{v}_{y, \mathrm{joint}}}
\newcommand{\vxcross}{\hat{v}_{x, \mathrm{cross}}}
\newcommand{\vycross}{\hat{v}_{y, \mathrm{cross}}}
\newcommand{\vxmarg}{\hat{v}_{x,{\rm self}}}
\newcommand{\vymarg}{\hat{v}_{y,{\rm self}}}

\newcommand{\zjol}{|\vzjoint \cdot \vz |}
\newcommand{\xjol}{|\vxjoint \cdot \vx |}
\newcommand{\yjol}{|\vyjoint \cdot \vy |}
\newcommand{\xcol}{|\vxcross \cdot \vx |}
\newcommand{\ycol}{|\vycross \cdot \vy |}
\newcommand{\xmol}{|\vxmarg \cdot \vx|}
\newcommand{\ymol}{|\vymarg \cdot \vy|}

\newcommand{\acrit}{a_{\mathrm{crit}}}
\newcommand{\bcrit}{b_{\mathrm{crit}}}
\newcommand{\ccrit}{c_{\mathrm{crit}}}
\newcommand{\thetacrit}{\theta_{\mathrm{crit}}}

\newcommand{\zmat}{\mathbf{Z}}
\newcommand{\rzmat}{\mathbf{R}_Z}

\newcommand{\identity}{\mathbbm{1}} 

\usepackage{array}
\newcommand{\blockspike}{{\setlength{\extrarowheight}{2pt}\begin{pmatrix}  a^2 \vx \vx^\top & a b\vx \vy^\top  \\ ab \vy \vx^\top & b^2 \vy \vy^\top\end{pmatrix} }}

\newcommand{\mysection}[1]{\section{#1}} 
\newcommand{\mysubsection}[1]{\subsection{#1}} 
\usepackage[T2A,T1]{fontenc}
\usepackage[utf8]{inputenc}
\usepackage[russian,english]{babel}

\begin{document}

\preprint{APS/123-QED}

\title{Better Together: Cross and Joint Covariances Enhance Signal Detectability in Undersampled  Data }

\author{Arabind Swain}
 \affiliation{Department of Physics, Emory University, Atlanta, GA 30322, USA}
\author{Sean Alexander Ridout}%
 \affiliation{Department of Physics, Emory University, Atlanta, GA 30322, USA}
 \affiliation{Initiative in Theory and Modeling of Living Systems, Atlanta, GA 30322, USA}
\author{Ilya Nemenman}
\affiliation{Department of Physics, Emory University, Atlanta, GA 30322, USA}
\affiliation{Department of Biology, Emory University, Atlanta, GA 30322, USA}
\affiliation{Initiative in Theory and Modeling of Living Systems, Atlanta, GA 30322, USA}

\date{\today}
\begin{abstract}
 Many data-science applications involve detecting a shared signal between two high-dimensional variables.  Using random matrix theory methods, we determine when such signal can be detected and reconstructed from sample correlations, despite the background of sampling noise induced correlations. We consider three different covariance matrices constructed from two high-dimensional variables: their individual self covariance, their cross covariance, and the self covariance of the concatenated (joint) variable, which incorporates the self and the cross correlation blocks.  We observe the expected Baik, Ben Arous, and P\'ech\'e detectability phase transition in all these covariance matrices, and we show that joint and cross covariance matrices always reconstruct the shared signal earlier than the self covariances. Whether the joint or the cross approach is better depends on the mismatch of dimensionalities between the variables. We discuss what these observations mean for choosing the right method for detecting linear correlations in data and how these findings may generalize to nonlinear statistical dependencies.
\end{abstract}
\maketitle

\mysection{Introduction}

Modern experiments measure increasingly large numbers of variables simultaneously, giving rise to extraordinarily large datasets. Examples include recordings from populations of neurons~\cite{cachannel1,neuropixels}, movies of animal postures~\cite{stephens2008dimensionality,Berman_etal_2014}, `omics datasets ~\cite{genesequence1,omics}, collective behavior~\cite{insectswarm}, ecological data~\cite{Dell_etal_2014}, etc. In many of these cases, one wants to understand the relationship between two high-dimensional variables---e.g., neural activity and behavior, or gene expression and cellular phenotypes. Such relations can be discovered by calculating matrices of various empirical linear correlations within and between the variables and finding the singular values and vectors of these correlation matrices. This is usually formalized via principal component analysis and regression (PCA and PCR), partial least squares (PLS), canonical correlation analysis (CCA), and other methods~\cite{wold1966estimation,Massy01031965,PCA1933}. 

In order to determine whether a specific singular value in a covariance matrix corresponds to a true signal or merely to sampling fluctuations, one typically starts by using random matrix theory (RMT) methods \cite{Potters_Bouchaud_2020} to calculate spectra of correlation matrices emerging from finite sampling effects in asymptotically uncorrelated data. These spectra are known for the self covariances \cite{marchenko1967распределение,Potters_Bouchaud_2020} and cross covariances~\cite{Forrester_2014,dupic2014spectraldensityproductswishart,Philiplinear, Rocks2022,PhysRevE2010ref,Potters_Bouchaud_2020} within and between high-dimensional variables. Roughly speaking, a  spectral outlier beyond this pure statistical noise is then statistically significant, and signals that produce such outliers can be estimated. Indeed, this intuition has been made precise for self covariances: when a signal magnitude crosses a certain threshold, the ability to detect the signal in the data self-covariance matrix undergoes a second order phase transition (the  Baik, Ben Arous, and P\'ech\'e, or BBP, transition \cite{BBP}), and the accuracy with which the principal vector corresponding to the largest eigenvalue  of the covariance matrix  characterizes the true signal rapidly increases from zero~\cite{BBP,BENAYCHGEORGES2011494,Potters_Bouchaud_2020}. Similarly, the asymptotic performance and limiting spectral distributions for high-dimensional CCA regime have been rigorously established~\cite{Bao2017CanonicalCC} along with the deviations between true and estimated signal~\cite{bykhovskaya2025highdimensionalcanonicalcorrelationanalysis}. To our knowledge, no definitive similar analysis exists for cross-covariances without whitening. In particular, it is not known how the ability of different linear methods to estimate low-rank correlations between two high-dimensional variables depends on properties of the variables, and numerical simulations suggest that this dependence is non-trivial \cite{abdelaleem2024simultaneous}. Our goal is to fill in this gap. A precise understanding of \emph{when} a low-rank correlation between $X$ and $Y$ can be detected, and \emph{how accurately} it can be characterized, requires a model of the signal. One reasonable model with a single signal is the \emph{latent feature model} (see, e.g., \cite{abdelaleem2024simultaneous,martini2024data,Philiplinear}):
\begin{align}
\xmat &= \rxmat + a \latentmat \vx^\top \label{eq:Xlatent}\\
\ymat &= \rymat + b \latentmat \vy^\top.\label{eq:Ylatent}
\end{align} 
Each row of the $T \times N_X$ ($T \times N_Y$) matrix $\xmat$ ($\ymat$) represents a sample from $X$ ($Y$). $\rxmat$ and $\rymat$ are uncorrelated Gaussian noise, with unit variance (generalization to $\sigma_X\neq 1$ and $\sigma_Y\neq 1$ is trivial, so that the unit variance assumption does not result in a loss of generality). True correlations between $X$ and $Y$ are encoded in $\mathbf{u}$, which contains $T$ independent samples of a one-dimensional ``latent'' variable $u$ with unit variance.  $\mathbf{u}$ is a $T\times1$ vector, with each component i.i.d. $\sim \mathcal N(0,1)$. This latent variable manifests itself in correlated signals, of variance $a^2$ and $b^2$, respectively, along directions given by the unit-norm vectors $\vx$ in $X$ and $\vy$ in $Y$. This model may be straightforwardly generalized to one with $r$ shared signals instead of one.

In this latent feature model, the concatenated variable $Z = (X,Y)$ is a sum of multivariate normals and thus has a normal distribution, with mean zero and covariance  
\begin{equation}
\Sigma = \identity  + \blockspike.
\end{equation}
Thus, $T$ samples from $Z$ can be generated from the standard white normal $\hat{\mathbf{Z}}$ through
\begin{equation}
\zmat = \hat{\mathbf{Z}} \sqrt{\Sigma}. \label{eq:genZ}
\end{equation}

Our goal is to simultaneously study three classes of methods. Firstly, we study methods which analyze the singular value decomposition (SVD) of the data matrices $\mathbf{X}$ and $\mathbf{Y}$ (e.g., PCA)---by definition, these singular values are the eigenvalues of the \emph{self-covariance} matrices $\mathbf{C}_{X} \equiv \frac{1}{T} \xmat^\top \xmat$ (and similarly for $Y$). Secondly, we consider methods which use the SVD of the \emph{cross-covariance} matrix, $\mathbf{C}_{XY} = \frac{1}{T} \xmat^\top \ymat$. Finally, we consider the detection of a signal using the \emph{joint-covariance} matrix, $\mathbf{C}_{Z} \equiv \frac{1}{T} \zmat^\top \zmat$. PLS, especially its Singular value decomposition (SVD) variant~\cite{PLSSVD}, works by performing SVD on the cross-covariance matrix ($X^{\top}Y$) between predictor variables and response variables. CCA, in contrast, uses the eigendecomposition of the \emph{whitened} cross-covariance, which is transformed using inverses of the $X$ and $Y$ covariance matrices, and is thus only possible when $T > N_X, N_Y$~\cite{Hotteling}. As we are interested in the under-sampled regime, we ignore CCA. All three analyses Joint PCA, PCA and PLS can be generated from a model of $\mathbf{C}_Z$, since

\begin{equation}
\mathbf{C}_Z = \begin{pmatrix}  \mathbf{C}_X  & \mathbf{C}_{XY}  \\ \mathbf{C}_{XY}^\top & \mathbf{C}_Y\end{pmatrix}.
\end{equation}

Equation \ref{eq:genZ} means that covariance and cross-covariance matrices are described by  \emph{multiplicative spike models} \cite{BENAYCHGEORGES2011494,Potters_Bouchaud_2020, johnstone2001, BBP} (``spike'' here is used for a low-rank deterministic perturbation to otherwise uncorrelated data).  In particular, the multiplicative spike model for the empirical joint-covariance matrix is 
\begin{equation}\label{Eq.multiplicativejointspike}
\mathbf{C}_Z = \frac{1}{T}\sqrt{\Sigma}  \hat{\zmat}^\top \hat{\zmat} \sqrt{\Sigma} \sim \frac{1}{T} \hat{\zmat} \mleft[ \identity  + \blockspike \mright] \hat{\zmat}^\top,
\end{equation}
where  $\sim$ denotes equality of the nonzero eigenvalues.  

Without the special structure introduced by distinguishing $X$ and $Y$, this and related models have been investigated repeatedly~\cite{Potters_Bouchaud_2020,Philiplinear,DING2021, DING2023,itamar2023}. As mentioned above, the self-covariance  matrix exhibits the \emph{BBP phase transition}, where the signal changes from undetectable to detectable at some threshold magnitude. Existing analytical results allow for the spectra of the joint-covariance matrix, and the self-covariance matrices $\mathbf{C}_{X} \equiv \frac{1}{T} \xmat^\top \xmat$ (and similarly for $Y$) to be computed. 

We are not aware of similar analytical results for the cross-covariance matrix, $\mathbf{C}_{XY} = \frac{1}{T} \xmat^\top \ymat$. In particular, the spectrum of $\mathbf{C}_{XY}$ \emph{cannot} be computed using the spectrum of $\mathbf{C}_Z$ alone. However, such analysis is necessary to compare the ability of cross-covariance based methods, like PLS, to methods which use the full covariance matrix. Thus, we introduce an \emph{additive spike model} of the joint-covariance matrix, which will allow this comparison to be made using existing techniques~\cite{BENAYCHGEORGES2011494,BENAYCHGEORGES2012120,Potters_Bouchaud_2020}. We will then verify numerically that our qualitative conclusions hold in the latent feature model as well.

Collecting the vectors $a \vx$ and $b \vy$ into a vector $c \vz$, the exact (sample) joint covariance of the latent feature model is
\begin{equation}
\mathbf{C}_Z = \frac{1}{T}\rzmat^\top \rzmat + \frac{c}{T} (\rzmat^\top \mathbf{u} \vz^\top + \vz \mathbf{u}^\top \rzmat) + \frac{c}{T} \mathbf{u}^\top \mathbf{u} \vz \vz^\top,
\label{eq:czexpand}
\end{equation}
where $\rzmat$ is the noise matrix formed by the concatenation of $\rxmat$ and $\rymat$. For a large number of samples, $\mathbf{u}^\top \mathbf{u} \approx T$. The cross-terms, further, are expected to have a small effect, because $\mathbf{u}$ and $\rzmat$ are uncorrelated. Thus, we expect the joint-covariance matrix to be {\em approximately} described by the \emph{additive spike model}
\begin{equation}
\mathbf{C}_Z= \frac{1}{T}\rzmat^\top \rzmat  + \blockspike. \label{eq:add}
\end{equation}
We do not expect this approximation to be quantitatively exact because the cross-terms in Eq.~(\ref{eq:czexpand}) are statistically dependent on $\rzmat^\top \rzmat $ and cannot  be neglected summarily. Additive spike models, however, show qualitatively similar phenomena to multiplicative spike models, such as the BBP phase transition \cite{BENAYCHGEORGES2011494}. Indeed, for a single variable $X$, the biggest distinction between additive and multiplicative spike models is a change in the spike magnitude at which the transition happens \cite{BENAYCHGEORGES2011494}. Thus, we expect analysis of this additive model to produce qualitatively accurate conclusions.

Thus, here we study the problem of correlating low-dimensional structures in two high-dimensional datasets using the additive spike model defined by Eq.~(\ref{eq:add}). Within this additive spike model, we separately analyze the empirical covariance spectra of $X$, $Y$, and $Z$, as well as the spectrum of the empirical cross-covariance between $X$ and $Y$.  We show that linear ``simultaneous dimensionality reduction'' techniques \cite{abdelaleem2024simultaneous,martini2024data}, where correlated low-dimensional subspaces of $X$ and $Y$ are found concurrently (e.g., PLS or PCA on the variable $Z$), generally perform  better than ``independent dimensionality reduction'' via PCA on $X$ and $Y$, followed by regressing the two sets of significant principal components on each other (PCR). We further show that, surprisingly, there is a regime where the correlation between $X$ and $Y$ is easier to  detect using $\mathbf{X}^\top \mathbf{Y}$ \emph{alone}, disregarding the information contained in the self covariances $\mathbf{C}_X$ and $\mathbf{C}_Y$. 

We end with results of numerical simulations, which suggest that our qualitative findings  hold for the latent feature model, Eqs.~(\ref{eq:Xlatent},~\ref{eq:Ylatent}) as well.
In parallel with our work, other authors have recently solved this latent feature model analytically~\cite{lenka}. Their exact solution could be used to extend our analyses to this model, which is likely a better model of real data.

\mysection{Models}
We start by rewriting the model, Eq.~(\ref{eq:add}),
as
\begin{equation}
\mathbf{C}_Z= \mathbf{W}_Z  + \mleft({a^2+b^2}\mright)\vz \vz^\top, \label{eq:add1}
\end{equation}
where 
\begin{multline}
\mathbf{W}_Z= 
 \frac{1}{T}\mathbf{R}^\top_Z\mathbf{R}_Z=\frac{1}{T}\begin{bmatrix}
        \mathbf{R}_X^\top \mathbf{R}_X& \mathbf{R}_X^\top \mathbf{R}_Y\\
        \mathbf{R}_Y^\top\mathbf{R}_X&\mathbf{R}_Y^\top \mathbf{R}_Y
    \end{bmatrix}\\
    \equiv
    \begin{bmatrix}
        \mathbf{W}_X& \mathbf{W}_{XY}\\
        \mathbf{W}_{YX}&\mathbf{W}_Y
    \end{bmatrix}.
\end{multline}
is the Wishart matrix of the concatenated, joint variable $Z$, and 
\begin{equation}
\vz=\mleft(\frac{{a}}{{c}}\vx,\frac{{b}}{{c}}\vy\mright),\quad\quad {c}^2={a}^2+{b}^2\label{eq:zeta}
\end{equation}
is the unit magnitude vector in the direction of the spike in this joint variable. $\vx$, $\vy$ and $\vz$ are all unit norm vectors.

Inspecting Eqs.~(\ref{eq:add}-\ref{eq:zeta}), we observe that the covariance matrix $\mathbf{C}_Z$ in the additive spike model can be written as self- and cross-covariance blocks, with additive spikes of different magnitude added to each block. Thus, within the {\em additive spike joint covariance model}, defined in Eq.~(\ref{eq:add}), we can also calculate the (empirical) {\em self-covariance matrix} of $X$,
\begin{equation}\label{spikecovariancex}
    \mathbf{C}_X= \mathbf{W}_X+ {a^2} \vx \vx^\top,
\end{equation}
the (empirical) {\em self-covariance matrix} of $Y$,
\begin{equation}\label{spikecovarianceY}
    \mathbf{C}_Y= \mathbf{W}_Y+ {b^2} \vy \vy^\top,
\end{equation}
and the (empirical) {\em cross-covariance matrix}
\begin{equation}\label{spikecrossvovariance}
    \mathbf{C}_{XY}=\mathbf{C}_{YX}^\top = \mathbf{W}_{XY}+ {ab} \vx \vy^\top.
\end{equation}
Thus, we can compare the ability of each of these matrices, and the joint-covariance matrix itself, to detect a given shared signal in $X$ and $Y$ (spike).

To  explore different regimes, we define the aspect ratios of different parts of the data matrix:
\begin{align}
    \qx \equiv \nx/T, \; \qy\equiv \ny/T,\;p_X\equiv1/\qx, \; p_Y\equiv1/\qy,
\end{align}
and we always assume $T,N_X,N_Y\to \infty$. 
Small $q$s and small $p$s mean  over- and under-sampling, respectively.
While the spectral distributions of the self‐covariance matrices in Eqs.~(\ref{spikecovariancex},~\ref{spikecovarianceY}) are classical results ~\cite{wishart,BBP,BENAYCHGEORGES2011494,Potters_Bouchaud_2020}, obtaining the spectra of the joint covariance $\mathbf{C}_Z$ and of the cross‐covariance $\mathbf{C}_{XY}$ requires some work.  

Before proceeding, we first note that we define a spike as detectable if, with matrix sizes going to infinity at fixed $\qx,\qy$, with probability one it produces a spectral outlier whose empirical singular vector has a nonzero overlap with the true direction in $X$ or $Y$; i.e., it sticks out above the noise bulk.  However, an outlier in only one self covariance ($\mathbf{C}_X$ or $\mathbf{C}_Y$) signals structure in that variable alone and does not establish an $X$--$Y$ correlation.  We, therefore, call detection of a shared signal ``successful'' if and only if the outlier’s singular vector(s) overlaps simultaneously with both $\vx$ and $\vy$.

\mysection{Results}

\mysubsection{Additive spike self covariances} \label{subseq:self}
First, we review known results, which will allow us to compute the spectra both for the self- and joint-covariance matrices. These are textbook results, listed here for completeness only, and a reader can skip them if they know the literature well.

Consider an additive spike $a \vv$ on the background of any square random matrix $\mathbf{A}$,
\begin{equation}\label{eq:gen_Additive}
    \tilde{\mathbf A}={\mathbf A}+a^2\vv \vv^\top.
\end{equation}
If $\mathbf{A}$ has spectral support $\lambda \in[\lambda_-, \lambda_+]$, the spike is detectable as an outlier in the spectrum of $\tilde{\mathbf{A}}$ for large enough signal strengths, $a>a_{\rm crit}$. $\acrit$  can be found using the Stieltjes transform $\mathfrak{g}_{\mathbf A}$ of $\mathbf{A}$~\cite{Potters_Bouchaud_2020}, as
\begin{equation}
\label{eq:St-limit}
\acrit^2 = \frac{1}{\mathfrak{g}_{\mathbf A}(\lambda_+)}.
\end{equation}
This outlier eigenvalue is associated with an outlier eigenvector $\vmax$. As long as $a>\acrit$, $\vmax$ has nonzero overlap with the spike $\vv$. Its value can be computed using the $\mathcal{R}$ transform, defined as
 \begin{align}\label{eq:Rtransform_defn}
     \mathcal R_\mathbf{A}(z)=\mathcal B_\mathbf{A}(z)-1/z\,,
\end{align}
where the $\mathcal B$-transform is the functional inverse of the Stieltjes transform
\begin{align}
     \mathcal B_{\mathbf A}[\mathfrak g_{\mathbf A}(z)]=z.
\end{align}
The overlap of $\vmax$ with the spike can then be calculated from the derivative of the $\mathcal{R}$-transform~\cite{Potters_Bouchaud_2020} as
\begin{equation}\label{rtransformoverlap}
     |\vmax\cdot\vv|=\sqrt{1-\frac{1}{(a^2)^2}\mathcal{R}'\mleft(\frac{1}{a^2}\mright)}.
\end{equation}

In our model, the self-covariance matrices are Wishart matrices, Eq.~(\ref{spikecovariancex}). In this case, the Stieltjes transform is well known~\cite{marchenko1967распределение,Potters_Bouchaud_2020}:
\begin{equation}
    \mathfrak g_{\mathbf{W}_X}(z)=\frac{z-1+\qx-\sqrt{z-\lambda_+}\sqrt{z-\lambda_-}}{2 \qx z},
\end{equation}
where $\lambda_{\pm}=(1\pm \sqrt{\qx})^2$. 
Thus, for the spike to produce a detectable outlier in the spectrum of the $X$ self covariance, one must have
\begin{equation}\label{condcovspikex}
    {a^2}\geq a^2_{\rm crit}= \frac{1}{\mathfrak g_{\mathbf{W}_X}(\lambda_+)}=\sqrt{\qx}\mleft(1+\sqrt{\qx}\mright).
\end{equation}
Using $\vxmarg$ to denote the eigenvector associated with this eigenvalue, its overlap with the true signal direction is then
\begin{equation}
     \xmol =\left\{\begin{array}{cc}
        \sqrt{1-\frac{\qx}{(a^2-\qx)^2}} & \text{if }  {a^2}\geq a^2_{\rm crit}, \\
         0 & \text{if }  {a^2}<a^2_{\rm crit}.
     \end{array}
     \right.
\end{equation}
Similarly, to detect an outlier in the $Y$ self covariance, one must have
\begin{equation}\label{condcovspikey}
    b^2 \geq \bcrit^2= \frac{1}{\mathfrak g_{\mathbf{W}_Y}(\lambda_+)}=\sqrt{\qy}\mleft(1+\sqrt{\qy}\mright),
\end{equation}
and the $Y$ spike direction is estimated with overlap
\begin{equation} \label{eq:y_overlap}
    \ymol=\left\{\begin{array}{cc}
        \sqrt{1-\frac{\qy}{(b^2-\qy)^2}} & \text{if }  {b^2}\geq b^2_{\rm crit}, \\
         0 & \text{if }  {b^2}<b^2_{\rm crit}.
     \end{array}
     \right.
\end{equation}

Overall, when analyzing the self-covariance matrices $\mathbf{C}_X$, $\mathbf{C}_Y$, the outlier eigenvectors will have a nonzero overlap with {\em both} the $X$ and the $Y$ components of the spike when both  conditions, Eqs.~(\ref{condcovspikex}, \ref{condcovspikey}) are satisfied {\em simultaneously}.

\mysubsection{Additive spike joint covariance}
The joint covariance spiked model is defined in Eq.~(\ref{eq:add1}). $\mathbf{W}_Z$ is still a Wishart matrix, regardless of our interpretation of the $X$ and $Y$ blocks as representing different observables. Thus, similarly to Subsection \ref{subseq:self}, an outlier can be detected in the spectrum of the joint covariance in the limit of very large matrix sizes if
\begin{equation} 
    c^2={a^2}+{b^2}\geq \ccrit^2=\sqrt{\qx+\qy}\mleft(1+\sqrt{\qx+\qy}\mright).\label{eq:joint_outlier}
\end{equation}
Further, the overlap of the eigenvector $\vzjoint$ associated with this outlier eigenvalue with the spike is
 \begin{equation}\label{eq:joint_overlap}
     \zjol=\left\{\begin{array}{cc}
        \sqrt{1-\frac{\qx+\qy}{(a^2+b^2-\qx-\qy)^2}} & \text{if }  c^2\geq \ccrit^2, \\
         0 & \text{if }  c^2< \ccrit^2.
     \end{array}
     \right.
\end{equation}

\begin{figure}
    \centering
    \includegraphics{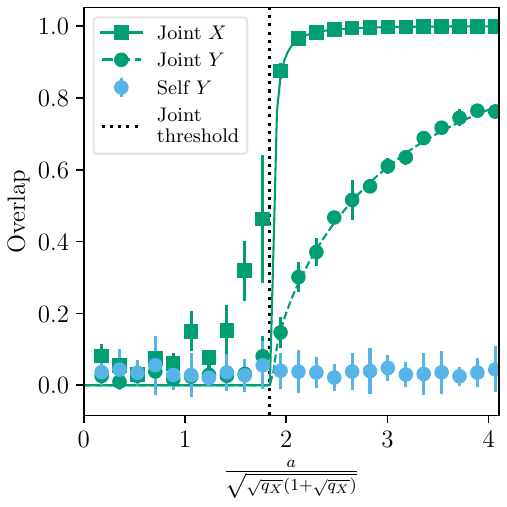}
    \caption{\textbf{Estimation of $X$ and $Y$ signals using the joint covariance.} We fix $b=0.5, \qx=1, \qy=4$ ($T=200$, $\nx=200$, $\ny = 800$), such that $b < \bcrit$, and then vary the $X$ signal strength $a$. As $a$  increases, in numerical simulations, both the $X$ (green squares) and $Y$ (green circles) components of the estimated spike $\vzjoint$ develop nonzero overlap with the true spike when $a^2+b^2$ crosses the threshold $\ccrit$ (Eq.~\ref{eq:joint_outlier}). Lines show analytical predictions, Eqs.~(\ref{eq:joint_overlap}, \ref{eq:joint_x}), which agree with numerical simulations, save for finite-size fluctuations. In contrast, $\vymarg$ always has zero overlap with the signal in $Y$, cf.~Eq.~(\ref{eq:y_overlap}) (blue circles).  Averaging is over $n=10$ independent simulations. Error bars are standard deviations.}
    \label{fig:concatoverlap}
\end{figure}

Recall that our criterion for success is nonzero overlap with  \emph{both} $\vx$ and $\vy$. Thus, we must check if detection of the outlier eigenvalue in $\mathbf{Z}$ guarantees that {\em both} self outlier directions $\vx$ and $\vy$ are correctly identified.  To answer this, we define the joint estimators of $\vx$ and $\vy$, $\vxjoint$ and $\vyjoint$, by projecting $\vzjoint$ into the $X$ or $Y$ subspaces and then normalizing the results. We call the quantity $\xjol$ the  joint $X$ overlap, and we similarly define the joint $Y$ overlap. 

A straightforward calculation (Appendix \ref{app:projection}), using only  axial symmetry and the limit $N_X, N_Y, T \to\infty$,  relates $\xjol$ to $\zjol$, 
\begin{equation} \label{eq:joint_x}
\xjol^2 = \frac{ 1 }{ 1 + (\left|\vzjoint \cdot \vz \right|^{-2}-1) \frac{\qx}{\qx + \qy} \frac{b^2+a^2}{a^2}},
\end{equation}
and similarly for $Y$. Together with Eq.~(\ref{eq:joint_overlap}), this shows that, whenever $c^2 > \ccrit^2$, both the joint $X$ overlap and the joint $Y$ overlap are nonzero.

Because $\sqrt{x} (1+\sqrt{x})$ is concave, $\ccrit^2 \leq \acrit^2 + \bcrit^2$. Thus, for any parameters where the correlation between $X$ and $Y$ can be detected using the two self-covariance matrices, it can \emph{also} be detected in the joint covariance (recall discussion under Eq.~(\ref{eq:y_overlap})). However, the converse is not true: there is a parameter regime when the spike \emph{cannot} be detected in one of the two self covariances, but it \emph{can} be detected in the joint covariance. 

We illustrate these findings in Fig.~\ref{fig:concatoverlap}, where we evaluate joint and self overlaps for  $N_Y > N_X = T$, so that at least $Y$ is severely undersampled. We keep  $b< \bcrit$ fixed, so that the spike {\em cannot} be detected in the $Y$ self-covariance ${\mathbf C}_Y$, and thus methods based on self covariances fail by our criterion.  We then vary the $X$ signal strength $a$. As expected, the self $Y$ overlap remains zero (within statistical fluctuations) for all $a$, and both joint $X$ overlap and joint $Y$ overlap undergo a second order phase transition {\em simultaneously} as $c$ crosses the $\ccrit$ threshold (detection below the threshold is possible due to finite-size fluctuations near the edge of the bulk spectrum~\cite{bloemendal2016principal}). 

\begin{figure}
    \centering
    \includegraphics{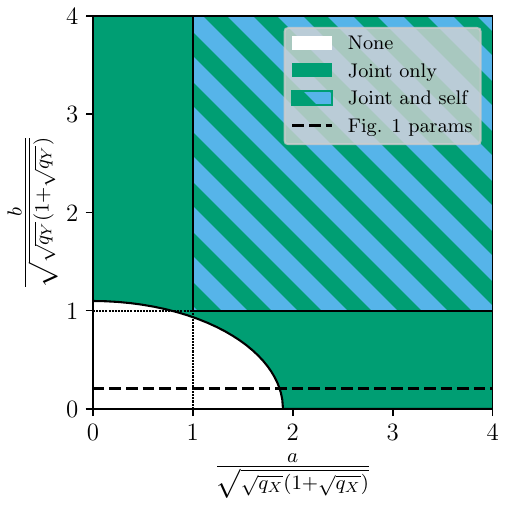}
    \caption[Phase Diagram for joint covariance and self individual covariances]{\textbf{Phase diagram for spike detectability from  self and joint covariances}. Solid green represents the region  where a spike results in a detectable outlier in the  joint-covariance matrix. In the region with alternate blue and green hatching, outliers are detectable by both methods. For the white region, none of the methods are able to detect a signal. For this plot $\qx=1$, $\qy=4$. The dotted lines give the bounds where a spike can be detected in the respective self-covariance. The dashed line represents the parameters used in Fig.~\ref{fig:concatoverlap}. }
    \label{Phaseconcat}
\end{figure}

We generalize these results and calculate the phase diagram for successful detection of a shared signal  for different values of $a$ and $b$, Fig.~\ref{Phaseconcat}, using Eqs.~(\ref{condcovspikex}, \ref{condcovspikey}, \ref{eq:joint_outlier}). The phase diagram has three regions. First, when both the $X$ and the $Y$ components of the spike signal are small, so that $c<\ccrit$ (white area), correct identification of the spike is impossible from either the self covariances (${\mathbf C}_X$ and  ${\mathbf C}_Y$) or the joint covariance ${\mathbf C}_Z$. Second, when the spike is sufficiently large in just the $X$ or the $Y$ subspace, $X$--$Y$ correlations can be successfully detected from projections of the joint eigenvector with the largest eigenvalue (green area). Yet, the signal cannot be detected in at least one (and sometimes both) subspaces from self covariances alone. Finally, when both $a$ and $b$ are large enough (blue and green hatching), detection is possible from either self (blue) or joint (green) covariances. Crucially, there does not exist a regime where detection via self covariances beats that via joint covariance.

\subsection{Spiked cross covariance model}
We will take advantage of existing results for a rectangular matrix with a spike~\cite{BENAYCHGEORGES2012120,itamar2023, pourkamali2024rectangularrotationalinvariantestimator} in order to compute the conditions for detection of a signal in the cross-covariance matrix. First, we define a general spiked rectangular matrix model as (compare to Eqs.~(\ref{spikecrossvovariance}, \ref{eq:gen_Additive}))
\begin{equation}\label{eq:recspike}\tilde{\mathbf{B}}=\mathbf{B}+\theta\vx \vy^\top .
\end{equation}
Here $\mathbf{B}$ is a $\nx\times\ny$ dimensional matrix, which has a singular value spectral support for $\lambda\in [\lambda_-,\lambda_+]$, and  $\vx$ and $\vy$ are $1\times \nx$  and $1\times\ny$ dimensional unit vectors, respectively. 
A method for computing the spectral outliers of such a model was proposed in Ref.~\cite{BENAYCHGEORGES2012120}. As in  the square-matrix problem (\ref{subseq:self}), there is a similar BBP transition, where an outlier appears when $\theta$ exceeds a threshold $\thetacrit$. But  different transforms and their inverses must be used for calculations. Specifically, one uses the $\mathcal{D}$-transform, 
\begin{equation}\label{eq:D transformdef}
    \mathcal{D}_{\mathbf {B}}(z)=z\mathfrak{g}_{\mathbf{B}\mathbf{B}^\top}(z^2)z\mathfrak{g}_{\mathbf{B}^\top \mathbf{B}}(z^2),
\end{equation}
which is related to the Stieltjes transform of the square matrix ${\mathbf B}^\top{\mathbf B}$, so the machinery used here is actually quite similar to the square case. The detectability threshold is then~\cite{BENAYCHGEORGES2012120}
\begin{equation} \label{eq:cross_crit}
{\thetacrit^2}   =\frac{1}{\mathcal{D}_{\mathbf{B}}(\lambda_+)}.
\end{equation}

Paralleling Sec.~\ref{subseq:self}, we define   $\mathbb D_{\mathbf{B}}$ as the functional inverse of the $\mathcal D$-transform. We further define $\lambda_{\rm max}$ as the expected maximum (outlier) singular value in $\tilde{\mathbf{B}}$ \cite{BENAYCHGEORGES2012120}, 
\begin{equation} \label{Eq:condlambda1} 
\lambda_{\rm max}=\left\{\begin{array}{cc} \lambda_+ & \text{if } {\theta}< {\thetacrit}, \\ \mathbb D_{\mathbf{B}}(\frac{1}{{\theta}^2}) & \text{if } {\theta}\geq {\thetacrit}. \end{array} \right. 
\end{equation} 
Then the expected overlaps between the left, $\vmax^{(l)}$, and the right,  $\vmax^{(r)}$, singular vectors corresponding to $\lambda_{\rm max}$ and the spike vectors $\vx$ and $\vy$ are \cite{BENAYCHGEORGES2012120}
\begin{align}
    |\vmax^{(l)}\cdot\vx|^2&=\left\{\begin{array}{cc}
        0 &  \text{if } {\theta}< {\thetacrit}, \\
       \frac{-2\lambda_{\rm max} \mathfrak{g}_{\mathbf{B}\mathbf{B}^\top}(\lambda_{\rm max}^2)}{{\theta}^2\mathcal{D}'_{\mathbf{B}}(\lambda_{\rm max})} & \text{if }{\theta}\geq {\thetacrit},
    \end{array}
    \right.
\\    |\vmax^{(r)}\cdot \vy|^2&=\left\{\begin{array}{cc}
        0 &  \text{if } {\theta}< {\thetacrit}, \\
       \frac{-2\lambda_{\rm max} \mathfrak{g}_{\mathbf{B}^\top\mathbf{B}}(\lambda_{\rm max}^2)}{{\theta}^2\mathcal{D}'_{\mathbf{B}}(\lambda_{\rm max})} & \text {if }{\theta}\geq {\thetacrit}.
    \end{array}
    \right.
\end{align}

To use these results in the special case of the cross-covariance matrix, when $\mathbf{B}=\mathbf{W}_{XY}$ and $\theta=ab$, as in Eq.~(\ref{spikecrossvovariance}), we need to evaluate  $\mathcal{D}_{\mathbf{W}_{XY}}$ and  $\mathbb{D}_{\mathbf{W}_{XY}}$. For this, we use the result for the Stieltjes transform of $\mathbf{W}^\top \mathbf{W}$ from~\cite{swain2025distributionsingularvalueslarge}, which calculates the bulk spectrum of the cross-covariance without any spikes. After some algebra, the result simplifies to:
\begin{multline}
    \mathcal{D}_{\mathbf{W}_{XY}}(z)=z\mathfrak{g}_{\mathbf{W}_{XY}\mathbf{W}_{XY}^\top}(z^2)z
    \mathfrak{g}_{\mathbf{W}_{XY}^\top\mathbf{W}_{XY}}(z^2)\\
    =\mleft(p_X z\mathfrak g_{0}(z^2)+\frac{1-p_X}{z}\mright)\mleft(p_Y z\mathfrak{g}_0(z^2)+\frac{1-p_Y}{z}\mright)\label{Dcrosscov},
\end{multline}
where  the terms proportional to $1/z$ in both parentheses come from zero singular values in  $\mathbf{X}$ and $\mathbf{Y}$, and $\mathfrak{g}_0$ is the Stieltjes transform corresponding to nonzero singular values only. $\mathfrak{g}_0$ does not have a simple analytical expression, but it satisfies the following equation \cite{swain2025distributionsingularvalueslarge}
\begin{equation}\label{eq:cubicstilt1}
    \alpha_3 \mathfrak g_0(z)^3+  \alpha_2 \mathfrak g_0(z)^2+ \alpha_1 \mathfrak g_0(z)+  \alpha_0=0,
 \end{equation}
where
\begin{align}
    &\alpha_3=z^2p_Xp_Y,\\
    &\alpha_2=z\mleft(p_Y(1-p_X)+p_X(1-p_Y)\mright),\\
    &\alpha_1=\mleft((1-p_X)(1-p_Y)-z p_Xp_Y\mright),\\
    &\alpha_0=p_Xp_Y.
\end{align}

We now define $\mathfrak{f}(z) \equiv z \mathfrak{g}_0(z^2)$ (cf. Eq.~(\ref{Dcrosscov})). This results in
\begin{equation} \label{eq:cubic_f}
    \alpha_3' \mathfrak{f}(z)^3+  \alpha_2' \mathfrak{f}(z)^2+ \alpha_1' \mathfrak{f}(z)+  \alpha_0'=0,
 \end{equation}
 where
\begin{align}
    \alpha_3'&=z^2 \px\py,\\
    \alpha_2'&=z \mleft(\py(1-\px)+ \px(1-\py)\mright),\\
    \alpha_1'&=\mleft((1-\px)(1-\py)-z^2 \px\py\mright),\\
    \alpha_0' &=z \px \py.
\end{align}

We can proceed in two ways. Firstly, we can obtain a ``semi-analytical'' solution for any parameter values by numerical solution of these equations. Secondly, we can obtain analytical solutions in a simplifying limit.
To obtain the semi-analytical solution, we solve this polynomial equation numerically, get the $\mathcal D$-transform from Eq.~(\ref{Dcrosscov}) and approximate its derivative using finite differences. Defining $\vxcross$ and $\vycross$ as the left and the right singular vectors corresponding to the largest singular value, we then get  for $ab >\sqrt{\frac{1}{ \mathcal{D}_{{\mathbf W}_{XY}}(\lambda_+)}}$,
\begin{align}\label{overlap1}
    \xcol^2&=
      \frac{-2\mleft(p_X \mathfrak f(\lambda_{\rm max})+\frac{1-p_X}{\lambda_{\rm max}}\mright)}{a^2b^2 \mathcal{D}'_{\mathbf{W}_{XY}}(\lambda_{\rm max})},  \\
    \ycol^2&=
      \frac{-2\mleft(p_Y \mathfrak f(\lambda_{\rm max})+\frac{1-p_Y}{\lambda_{\rm max}}\mright)}{a^2b^2 \mathcal{D}'_{X_{X^T Y}}(\lambda_{\rm max})},  
\end{align}
and the overlaps are zero for smaller $ab$.

To obtain an analytical solution in a special case, we note that the spectral edge $\lambda_+$ for the singular value spectrum was found in \cite{swain2025distributionsingularvalueslarge}, and the expression is especially simple when  $p_Y = \epsilon p_X$, with $\epsilon \ll 1$, so that $\ny\gg\nx$. Specifically, in this case
\begin{align}
\lambda_{+}&\approx\sqrt{\frac{1+p_X+2\sqrt{p_X}}{p_Xp_Y}}.
\end{align}
Further, Eq.~(\ref{eq:cubic_f}) also simplifies in this case. Combining them, we get
\begin{equation}
    \mathfrak f (\lambda_+)\approx \sqrt{\py}.
\end{equation}
Then, with $\theta=ab$, the condition, Eq.~(\ref{eq:cross_crit}), to have an outlier with nonzero overlaps with the spike (that is, for analysis of the cross-covariance spectrum to be successful in detecting the signal)  transforms into
\begin{equation}
\label{eq:regioncross}
    ab\geq \thetacrit = \sqrt{\qy (\qx + \sqrt{\qx})}= \acrit \sqrt{\qy},
\end{equation}

To obtain a formula for the cross overlaps in this limit, we must first determine the outlier eigenvalue $\lambda_{\mathrm{max}}$. We know that $\lambda_+ \sim \sqrt{\qy}$ in this limit, so we expand the equation for $\mathcal{D}(\lambda_{\mathrm{max}})$ to lowest order in $\py$ under the assumption that $\lambda_{\mathrm{max}} = O( \sqrt{\qy})$. Plugging this into Eq.~(\ref{Eq:condlambda1}) and solving yields
\begin{equation}
\lambda_{\mathrm{max}} \approx \begin{cases}\lambda_+, & ab  \leq \thetacrit, \\
\lambda_+ \frac{a b}{\thetacrit} \sqrt{\frac{a^2 b^2 - \thetacrit^2 + \sqrt{\px} \thetacrit^2}{a^2 b^2 - \thetacrit^2 + \sqrt{\px} a^2 b^2}}, & ab >  \thetacrit.
\end{cases}
\end{equation}

Evaluating the lowest-order expressions for $\mathfrak{f}(\lambda_{\mathrm{max}} )$ and $\mathcal{D}'(\lambda_{\mathrm{max}} )$ (now assuming $a b = O(\sqrt{\qy})$) then gives
\begin{align}
\ycol^2 &\approx \begin{cases}1 - \frac{\px \thetacrit^2 a^2 b^2}{t_{\alpha}t_{\beta}}, & ab > \thetacrit,\\ 0, &a b \leq \thetacrit, \end{cases}\\
\xcol^2 &\approx \begin{cases}1 - \frac{\px \thetacrit^4}{t_{\alpha}^2}, & ab > \thetacrit,\\ 0, & ab \leq \thetacrit,\end{cases}
\end{align}
where $t_{\alpha}=\sqrt{\px} a^2 b^2 + a^2 b^2 -  \thetacrit^2$ and $t_{\beta}=\sqrt{\px} \thetacrit^2+ a^2 b^2 -  \thetacrit^2$.

\begin{figure}
    \centering
    \includegraphics{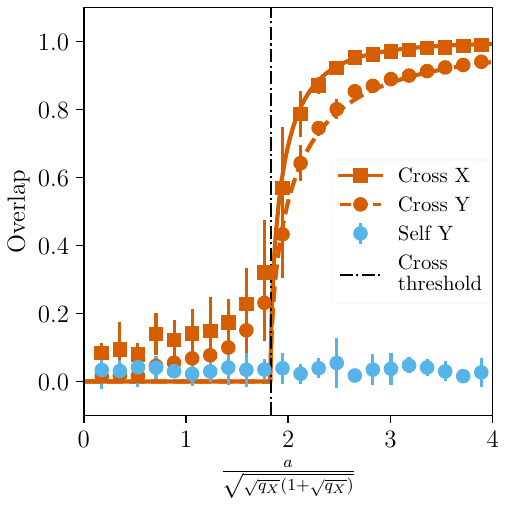}
    \caption{\textbf{Estimation of $X$ and $Y$ signals using the cross covariance.} We fix $b=2.5, \qx=1, \qy=20$ ($T = 100$, $\nx = 100$, $\ny = 2\times10^3$), such that $b < \bcrit$, and then vary the $X$ signal strength $a$. As $a$ is increased, in numerical simulations, both $\vxjoint$(orange squares) and $\vyjoint$ (orange circles) develop nonzero overlap with the true spike when $a b$ crosses the threshold, determined semi-analytically. Lines show semi-analytical predictions for the overlaps, which agree with numerical simulations, save for finite-size fluctuations. In contrast, $\vymarg$ always has zero overlap with the signal in $Y$, cf.~Eq.~(\ref{eq:y_overlap}) (blue circles).   Averaging is over $n=10$ independent simulations. Error bars are standard deviations.}
    \label{fig:overlapcross}
\end{figure}

In Fig.~\ref{fig:overlapcross}, we compare the semi-analytical cross overlaps to the empirical cross overlaps in simulated data. We also compare them to self overlaps, similar to Fig.~(\ref{fig:concatoverlap}). The agreement between the theory and the simulations is excellent again, showing a BBP-like detectability transition. Further, for these parameter values, it is clear that the cross-covariance matrix detects the spike well before {\em both} self-covariance matrices do.

\begin{figure}
    \centering
    \includegraphics{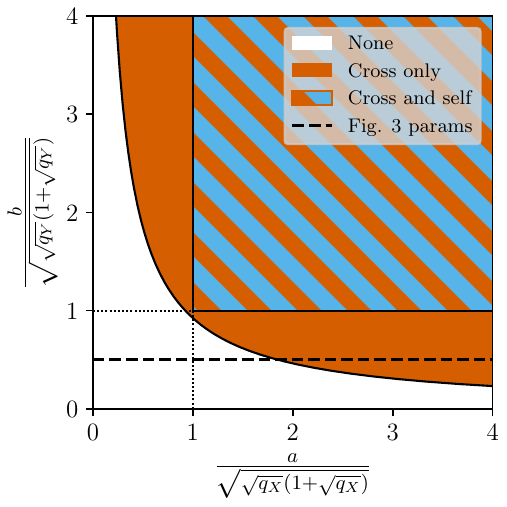}
    \caption[Phase Diagram for cross covariance and self]{\textbf{Phase diagram for spike detectability for cross  and self covariances.} We fix $\qx=1$, $\qy=20$ (notice that the value of $\qy$ is different from Fig.~\ref{Phaseconcat}, so that advantages of the cross-covariance approach are easier to see). We study how the signal strengths $a$ (for $X$) and $b$ (for $Y$) affect spike detection. In the red region, computed semi-analytically, both the $X$ and $Y$ components of the spike can be partially reconstructed (nonzero overlap).  The blue region is where the self covariances of $\mathbf{X}$ and $\mathbf{Y}$ can both detect their spikes, thus providing information about the entire spike. Thus, alternating blue and red stripes mark the region where both approaches give nonzero overlaps with the spike (though the magnitudes of the overlaps may be different). Crucially, the cross covariance may detect the spike when the self covariances cannot, but not the other way around. In the white solid region, neither method can detect the spike.}
    \label{phasecross}
\end{figure}

We formalize this superiority of the cross-covariance based detection  by exploring the phase diagram of the spike detectability as a function of the spike magnitudes,  $a$ and $b$, normalized such that the spikes in self-covariances can be detected at exactly 1.0 on both axes, Fig.~\ref{phasecross}. We consider a case where $\qx \ll \qy$, but construct the phase diagram using the exact Eq.~(\ref{eq:cross_crit}) (semi-analytically). We observe that, in the undersampled regime, when either $\qx\gg1$ or $\qy\gg 1$, the spike is always detectable in cross covariance before it can be detected in {\em both}  individual self covariances. As for the joint covariance (Fig.~\ref{Phaseconcat}), a strong spike component in the smaller-dimensional variable (here $X$), can make the weaker component in the larger-dimensional variable (here $Y$) easier to detect. Further, for some parameter combinations, the spike can be detected in the cross covariance when  \emph{neither} of the self covariances can detect it (to the left and below $[1,1]$ in the phase diagram). 

\mysubsection{Comparison between cross covariance and joint covariance}

The cross and joint covariance are superior to self covariances for detection of the spike in both variables. Here we analyze how these two methods compare to each other. To begin, we recall the general analytical result for the joint covariance spike detection threshold, Eq.~(\ref{eq:joint_outlier}), as well as the simplified analytical results for the cross-covariance detection threshold in the limit $\qy \gg \qx$, Eq.~(\ref{eq:regioncross}). To build intuition  and develop a simple heuristic for comparing spike detectability in both methods, we further simplify these results by focusing on the severely undersampled regime, $\qx,\qy\gg1$, which is common in modern data science. The spike detectability condition for the joint covariance becomes:
\begin{equation}
a^2 + b^2 \gtrsim  \qx + \qy \approx \acrit^2 + \bcrit^2,
\end{equation}
where $\acrit$ and $\bcrit$ are the thresholds for spike detection in the self covariance, Eqs.~(\ref{condcovspikex}, \ref{condcovspikey}). In contrast, when $\qy \gg \qx$ and $\qy \gg1$, the detectability condition for the cross covariance, Eq.~(\ref{eq:regioncross}), is
\begin{equation}
a b  \gtrsim \acrit \sqrt{\qy} \approx \acrit \bcrit.
\end{equation}

Recall that, by the AM--GM inequality, $x + y \geq 2 \sqrt{xy}$ for nonnegative $x$ and $y$. More importantly for us, the difference between the two is larger when $x$ and $y$ are more different. Thus, the criterion for the cross covariance will be easier to satisfy than the criterion for the joint covariance when $\qx \ll \qy$, but $a$ and $b$ are similar. On the other hand (although the approximation we have made for the cross covariance will not be valid), we expect that the joint covariance will work better when $a$ and $b$ are quite different, but $\qx$ and $\qy$ are similar. 

\begin{figure}
    \centering
    \includegraphics{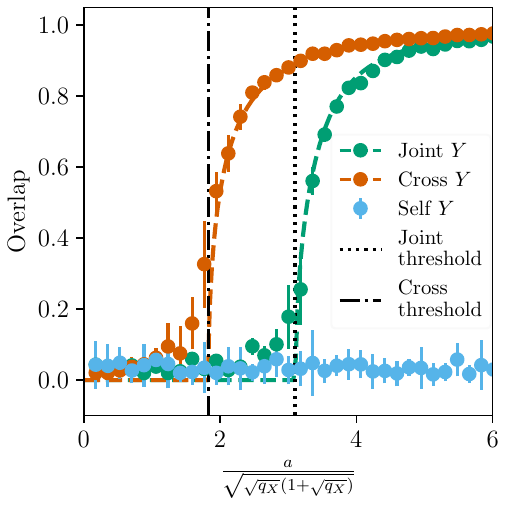}
    \caption[Comparison between joint and cross overlap]{\textbf{Comparison between joint and cross overlaps for estimating the spike in $Y$.} We fix $b=2.5, \qx=1, \qy=20$ ($T=100$, $\nx=100$, $\ny=2\times10^3$) such that $b < \bcrit$, and $\qy \gg \qx$, and then vary the $X$ signal strength $a$. As $a$ is increased, in numerical simulations, both $\vycross$ (orange circles) and $\vycross$ (green circles) develop nonzero overlap with the true spike $\vy$. Colored dashed lines show analytical (joint) and semi-analytical (cross) predictions. In this regime, where $Y$ is much more poorly sampled than $X$, there is a region where the cross $Y$ overlap is large, yet the joint $Y$ overlap is zero.  Dotted and dash-dotted black lines represent the analytically (or semi-analytically) calculated BBP transition values for the joint $Y$ overlap and cross $Y$ overlap, respectively.  Averaging is over $n=10$ independent simulations. Error bars are standard deviations.}
    \label{jointcrosscompare}
\end{figure}

\begin{figure}
    \centering
    \includegraphics{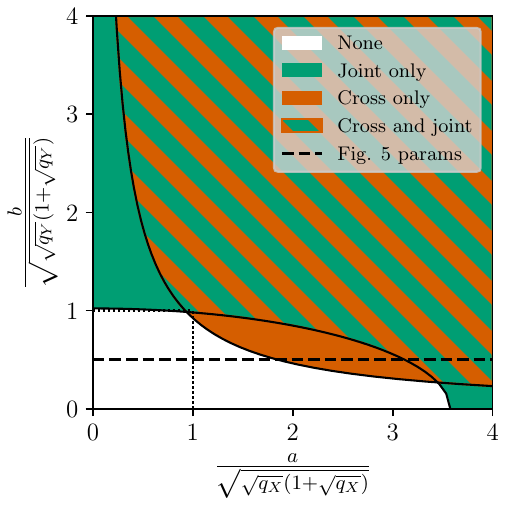}
    \caption[Phase Diagram for cross covariance and joint covariance]{\textbf{Phase diagram for spike detectability for joint  and cross covariances.} We fix $\qx=1$, $\qy=20$, and study how the signal strengths $a$ (for $X$) and $b$ (for $Y$) affect spike detectability. In the red region, computed semi-analytically, both the $X$ and $Y$ spikes can be partially reconstructed from the cross-covariance matrix (nonzero overlap). Green shows where both spikes can be partially reconstructed with the joint-covariance matrix. Thus, solid regions show where only one of the two methods is successful, while in the region with alternating green and red stripes, both approaches have nonzero overlaps with the spike (though the magnitudes of the overlaps may be different). In the white solid region, neither method can detect the spike. The dashed line shows the line of spike strength parameters used in Fig.~\ref{jointcrosscompare}}
    \label{jointcrosscompphase}
\end{figure}

Empirically, this heuristic works  well even when only one of the variables is undersampled. In Fig.~\ref{jointcrosscompare}, we compare the  $Y$ overlaps observed for different methods as a function of changing $a$ for a fixed $b$. $\qy\gg\qx,$ and $b$ are fixed to the same values as in Fig.~\ref{fig:overlapcross}, so that the spike in $Y$ cannot be detected in its self-covariance matrix. Crucially, for these parameters, the cross $Y$ overlap is larger than the joint one. This is because the example in the figure is in the limited area of the phase diagrams, Figs.~\ref{Phaseconcat}  and \ref{phasecross}, where an outlier in the cross covariance is expected to be easier to detect than in the joint covariance. We summarize this in Fig.~\ref{jointcrosscompphase}, where the phase diagrams of joint and cross covariance spike detection are compared.

That a region where cross covariance outperforms joint covariance exists is surprising, since the cross-covariance matrix is only a subset of the joint-covariance matrix. Naively, one would expect that, by incorporating more information, one should make spike detection easier, and thus the joint covariance  should never be inferior. Instead, we find that sometimes \emph{``throwing out''} the self parts of the joint-covariance matrix improves the inference! Intuitively,  this is because a very high-dimensional, undersampled self covariance block (e.g., for $\qy=20$) introduces a lot of opportunities for spurious correlations within the corresponding variable, $Y$. The increased dimensionality of the joint-covariance matrix compared to the cross-covariance one then outweighs the advantage provided by the data in the self-covariance block.

\section{Comparing cross covariance and self covariance in the  latent feature model}

\begin{figure}
    \centering
    \includegraphics{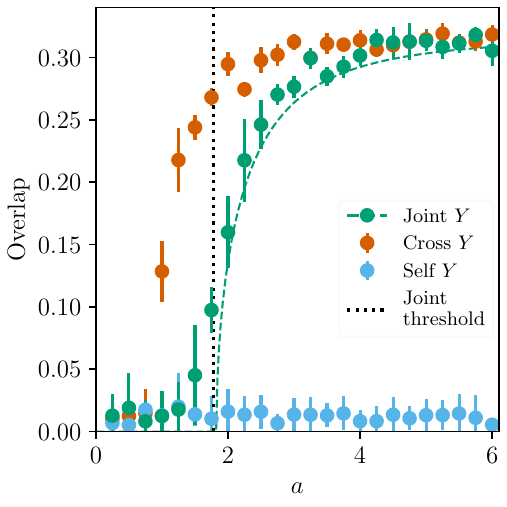}
    \caption[Comparison between joint and cross overlap for latent feature model]{\textbf{Comparison between joint and cross overlaps for the latent feature model.} We fix $b=1.5, \qx=1, \qy=20$ ($T=100$, $\nx=100$, $\ny=2\times10^3$)  such that $b < \bcrit$ and $\qy \gg \qx$, and then vary the $X$ signal strength $a$. As $a$ is increased, in numerical simulations, both $\vycross$ (orange circles) and $\vycross$ (green circles) develop nonzero overlap with the true spike $\vy$. As in the additive spike model (Fig.~\ref{jointcrosscompare}), the signal is detected in cross covariance for smaller values of $a$ than are required for the joint covariance. The dotted black line represents the analytically calculated BBP transition value for the Joint $Y$ overlap, and the green dashed line is the analytical prediction for the joint $Y$ overlap in this model (Appendix \ref{app:joint_latent}). Averaging is over $n=10$ independent simulations. Error bars are standard deviations.}
    \label{jointcrosscomparelatent}
\end{figure}

Since it seems counterintuitive that it is sometimes easier to detect a spike in the cross covariance than the joint covariance, we would like to confirm that this region in the phase diagram exists in other models, beyond the additive model considered here. For this, we investigate its existence in the  latent feature model, Eqs.~(\ref{eq:Xlatent}, \ref{eq:Ylatent}), numerically.
Figure~\ref{jointcrosscomparelatent} shows simulations of the latent feature model for parameters similar to the additive spike model in Fig.~\ref{jointcrosscompphase}. (Note that identical values of $a$ and $b$ are {\em not} equivalent in these models; the self-detection thresholds, for example, are different). For the joint case, analytical results can be obtained from existing work~\cite{BBP,wishart} (Appendix \ref{app:joint_latent}), by again using our calculations that convert the joint $Z$ overlap to the joint $Y$ overlap (Appendix \ref{app:projection}).  These simulations show that all our qualitative results are reproduced in the latent feature model. Firstly, for both the joint- and cross-covariance matrices, a strong enough signal in $X$ (large $a$) allows one to detect the direction of the spike in $Y$. Note, however, the difference in the extent of this effect: the joint and cross $Y$ overlaps plateau at a finite value as $a \to \infty$, rather than becoming 1 as in the additive model. Secondly, for $\qy \gg \qx$, the cross-covariance matrix again detects the signal in $Y$ more easily than the joint-covariance matrix.

Again, we note that others~\cite{lenka} have recently solved this model, and thus it should be possible to confirm these results analytically.

\section{Experimental test}

\subsection{Data: Bengalese finch song} 

We now test these ideas on experimental data. We study spectrograms of vocal gestures, or {\em syllables}, isolated from recordings of the song of adult Bengalese finches (see~\cite{tang2014millisecond} for description of the experiment). Each syllable spectrogram was constructed by binning time and then computing a Fourier transform of the spectrum within that time bin to assign a (log) power to a sequence of frequency bins (see \cite{tang2014millisecond} for details). The spectrograms were previously manually classified into different classes, labeled by the syllable type,  e.g., ``K'' or ``R''. It is known that spectral properties of sequential syllables are correlated \cite{wohlgemuth2010linked}, and we use this to construct a paired dataset to verify the ability of different linear methods to detect such dependencies.

Specifically, we identify each instance where a ``K'' syllable is immediately followed by an ``R'' in a single day's recording from a single finch, resulting in 318 such paired spectrograms. We further discard 14 outlier pairs where the $K$ spectrogram had an uncharacteristically low (below $0.8$) with the mean $K$ spectrogram, which we believe could have been misclassified in the original dataset. 

Syllables of even the same type vary in durations, but all three dimensionality reduction techniques considered here require fixing $N_X$ and $N_Y$. We thus linearly interpolate the spectrograms,  rescale the time axis to the same length as the longest syllable of each type, and re-bin the spectrograms along the time axis into 30 and 21 bins for K and R, respectively (in proportion to their average  duration). Both have 256 frequency bins. Thus, overall, our dataset contains $T_{\rm tot}=304$ samples  of paired spectrograms, with $X$ and $Y$ representing K and R syllable spectrograms, with  $N_X=256\times30=7710$ and $N_Y=256\times21=5397$.

\begin{figure*}
\includegraphics[width=\textwidth]{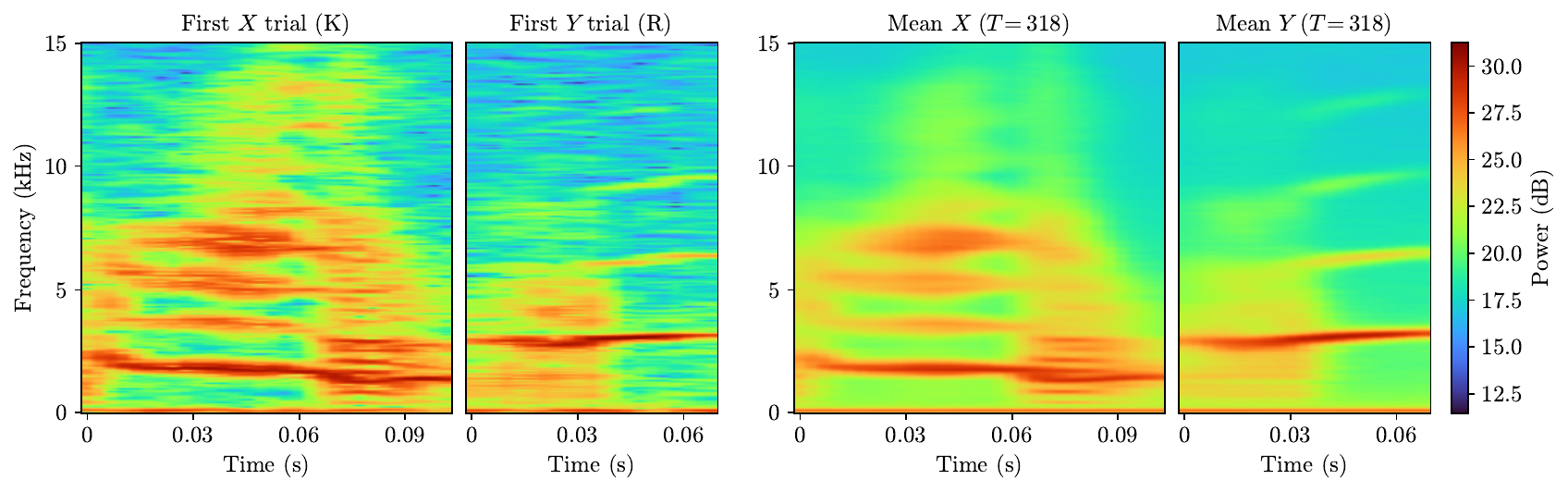}
\caption{Individual examples of preprocecessed K and R spectrograms (which are the $X$ and $Y$ variables in this example), as well as the mean spectrograms over all $T=318$ paired samples. Here $N_X=30$ and $N_Y=21$ bins.} \label{fig:spectrograms}
\end{figure*}

We expect the largest joint signal in the data to be simply volume: the distance between the bird and the microphone is not perfectly fixed. Since we expect distance from the microphone to act as a multiplicative rescaling of all powers, we subtract the mean log power from each syllable's spectrogram. An example of paired spectrograms, after all  preprocessing steps is shown in Fig.~\ref{fig:spectrograms}, alongside the mean spectrograms. 

Finally, we construct a second dataset where only $N_Y=10$ central time bins are included for $Y$, to try to test the prediction that decreasing $N_Y/N_X$ will improve the  performance of the cross-covariance method relative to other approaches. This is not a perfect test of our predictions, because the theoretical analysis assumed that the overall signal \emph{strength} was fixed for the changing $N_Y/N_X$ ration, whie  this ``trimming'' of the spectrogram will also changes the signal strength by an unknown amount. We hope, however, to still see an effect of the predicted sign.

With this preprocessed data, we apply the marginal, joint, and cross dimensionality reduction techniques in the standard way: rescaling each feature (spectrogram bin) by its standard deviation across the training set, and then computing the eigenvectors or singular vectors of the relevant data matrix.

\subsection{Results}
Unlike in our theoretical analysis, we cannot know in advance what the ``true'' signal is. This makes it difficult to identify precisely which method performs best on this experimental data. Nonetheless, we still hope to test our qualitative conclusions that SDR outperforms IDR for undersampled datasets, and that the cross-covariance method outperforms joint reduction when $N_X$ and $N_Y$ are very different.

\begin{figure}[t]
\includegraphics[width=\columnwidth]{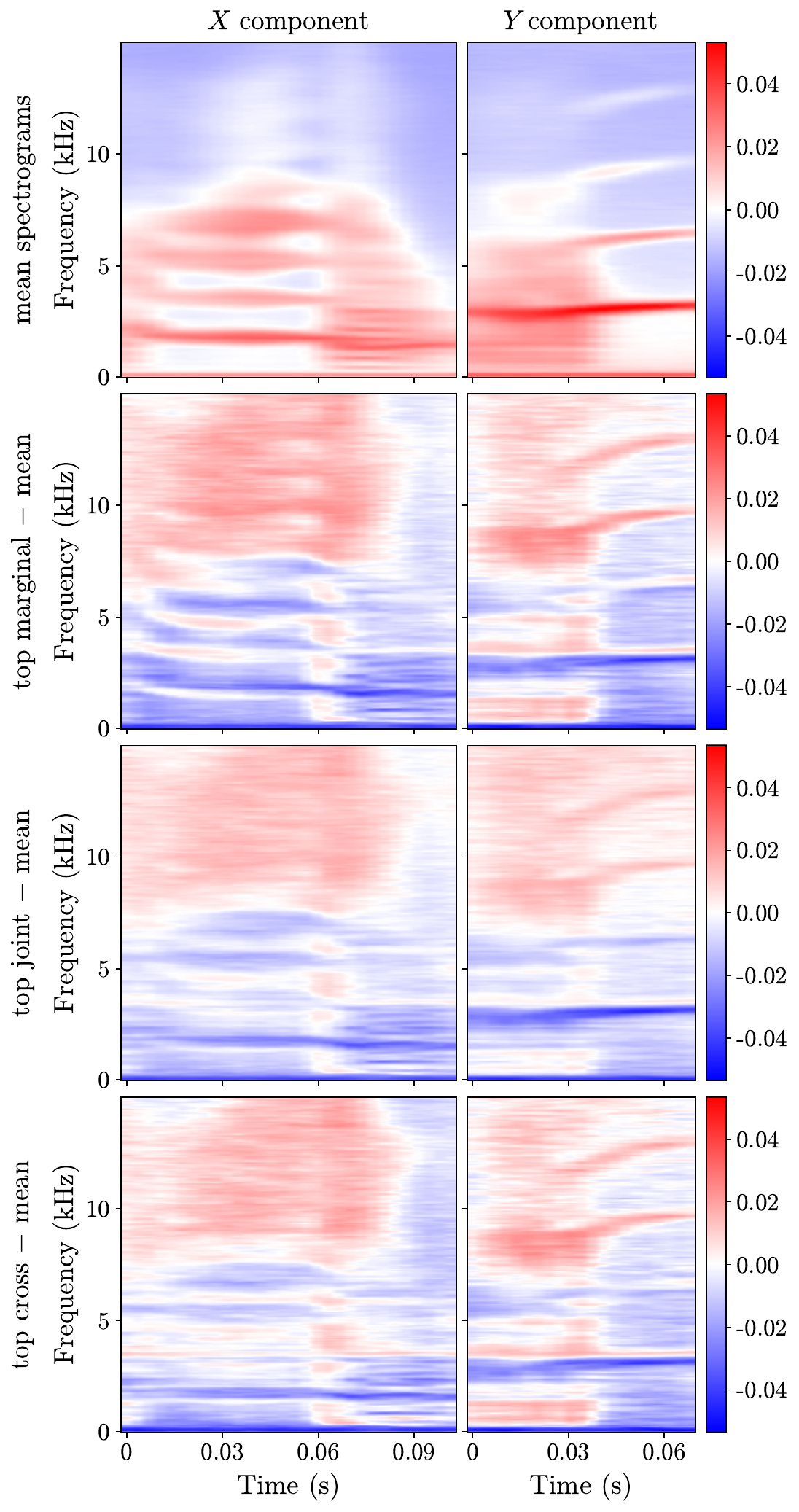}
\caption{(top) Normalized mean spectrograms of both syllables. (three bottom rows) Top eigenvectors/singular vectors associated with largest eigenvalues / singular values for each method are plotted after subtracting the normalized mean spectrograms. Note that in all cases, the top signals are very similar, and all signal higher power in the high frequency bins, suggesting that the joint signal being identified is a shared shift in the fundamental frequency of subsequent syllables.} \label{fig:loadings}
\end{figure}

Figure~\ref{fig:loadings} examines the top signals detected by all three methods: the top marginal eigenvectors for $X$ and $Y$, the normalized $X$ and $Y$ components of the top joint eigenvector, and the top left and right singular vector pair of the cross-covariance. To visualize what these signals are, in the first row we plot 
the mean spectrograms for $X$ and $Y$, which is similar to Fig.~\ref{fig:spectrograms}, but now evaluated without the outliers ($T_{\rm tot}=304$ samples), with each panel normalized to one. We then illustrate the top detected signals by the difference between the signal and these mean spectrograms. First, the the signals detected by all three methods for full data are very similar to each other, allowing us to use all three of them as proxies for the true signal (note that subsequent eigenvectors and singular vectors show a much higher variability across the methods). Secondly, the meaning of this top signal component is also clear: it detects higher power at high frequencies, including increase of the fundamental frequency of syllables. The latter  is clearly visible for the $Y$ panels, where the fundamental frequency band in the mean spectrogram is replaced by a pair of blue-red bands, so that the signal corresponds to observing the fundamental frequency in the upper part of its possible range. Such correlations among spectral properties of subsequent syllables are well-known \cite{wohlgemuth2010linked}. 

With this, we can now test the accuracy of each detection method in the undersampled regime relative to performance of all methods when well-sampled. To avoid train-test contamination, we first split our data randomly into 10 folds, and assign 9 folds to a ``large''  set (size $T_{\mathrm{large}}=0.9\times T_{\rm tot}$) and one fold to a `` small'' set (size $T_{\mathrm{small}}=0.1\times T_{\rm tot}$). For each of the 10 possible large/small splits, we apply each of the three methods to the large dataset to produce three possible proxies for the true signal, and to the small dataset to produce small-sample estimated signals.  For each $A \in {X, Y}$, $\alpha,\gamma \in \{\mathrm{marginal}, \mathrm{joint}, \mathrm{cross}\}$, we then ask how well the ``small-sample signal'' $\hat{v}_{A,\alpha,\mathrm{small}}$ is correlated with the ``proxy true signal'' $\hat{v}_{A,\gamma,\mathrm{large}}$, defining:
\begin{equation}
|r_{A,\alpha,\beta}| \equiv \hat{v}_{A,\alpha,\mathrm{small}}^\top  \hat{v}_{A,\gamma,\mathrm{large}}.
\end{equation}
For example, $r_{X,\mathrm{joint},\mathrm{marginal}}$ measures how well the small-sample estimated signal using the joint method correlates with the proxy for the ground-truth signal (large sample) obtained using the marginal method. 
Since all three methods produced fairly similar signals with large samples (Fig.~\ref{fig:loadings}), we expect that if method $\alpha$ truly has better small-sample performance than method $\beta$, we will find $|r_{A,\alpha,\gamma}|  \geq |r_{A,\beta,\gamma}|$ for most $A, \gamma$---even for $\gamma = \beta$. 

\begin{figure*}
\includegraphics[width=\textwidth]{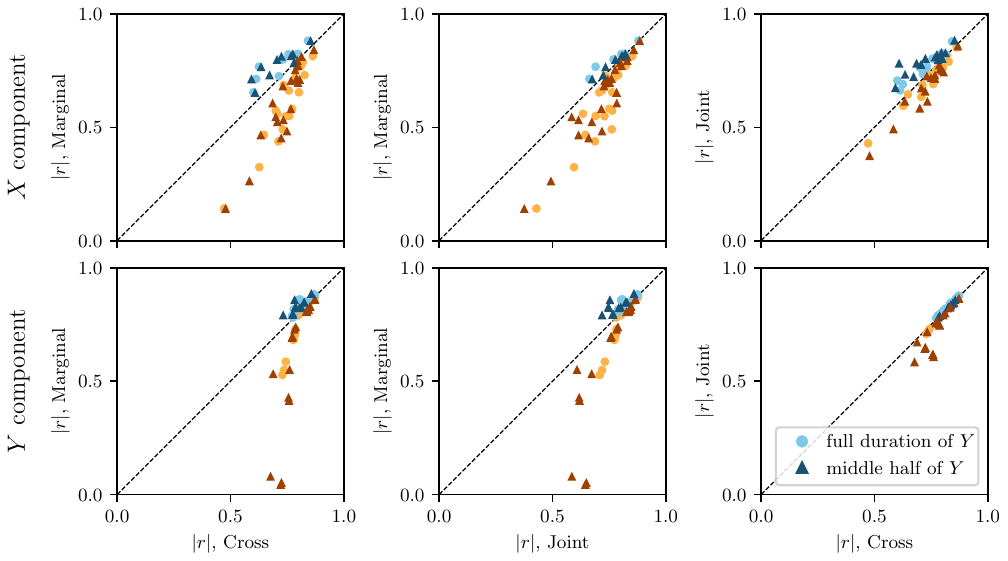}
\caption{Comparison of the correlation $|r|$ of the $X$ and $Y$ signals inferred from \emph{small} datasets with the putative ground-truth vectors inferred on \emph{large} datasets. The 30 points on each plot correspond to 10 different splits of the data and 3 different, nearly-equivalent choices of which method's large-sample result to identify the ``ground truth'' with. All three methods usually identify a signal close to the large-sample signal (cloud of points near $\left|r\right| \approx 0.8$. The marginal method, however, often fails, producing much smaller values of $\left|r\right|$. Circles show results for the original dataset, while triangl show results for a reduced-$N_Y$ dataset where half of the time bins are trimmed from the $Y$ spectrograms, keeping the middle $10$ bins. Notice that, especially on the trimmed dataset, the worst splits produce slightly worse results for the joint method than for the cross method, but this is a small effect.  } \label{fig:correlation}
\end{figure*}

Figure~\ref{fig:correlation} shows the result of this analysis for both the full data (light circles) and the dataset where $Y$ has been trimmed to its 10 central bins (dark triangles). All panels show $|r_{A,\alpha,\gamma}|$ vs.  $|r_{A,\beta,\gamma}|$, with all three possible choices of $\gamma$ pooled together and shown on the same plot. Top row is $A=X$ and bottom row is $A=Y$. Points are colored blue or orange based on whether the method indicated on the $y$ axis or the method indicated on the $x$ axis has a larger value of $|r|$.
While there are only two independent comparison combinations among the three methods, we admit some redundancy, and the three  columns in Fig.~\ref{fig:correlation} show all three pairwise method comparisons.  

Firstly, all panels show a large cloud of large-small splits with $|r| \sim 0.7$--$0.9$. For these random small samples, both methods work, and the small difference in accuracy between the two methods is arguably not meaningful, given the imprecise comparison we have been forced to make by our lack of ground-truth knowledge.  Secondly, many panels show a ``tail'' of low-accuracy results---for some small samples, one or both methods fails to identify a signal with large overlap with the proxies for the true signal. We observe that in all cases, this failure occurs for the \emph{marginal} estimator of the signal. Both joint methods consistently produced $|r| >0.5$. 

Further, in the joint v.\ cross (two rightmost) panels, we observe  that, although both methods essentially never dramatically failed, the lowest values of $r$ are slightly worse for the joint method (below the dashed line), especially when the dimensionality of $Y$ has been reduced. This is consistent with our theoretical predictions.

\section{Discussion}
We studied a set of additive spike models (which approximate the distribution of data under sampling noise and a shared signal) for joint covariance,  cross covariance and individual self covariances to understand when these matrices allow for detection of correlation between two high-dimensional variables $X$ and $Y$---that is, detection of eigenvectors or singular vectors with nonzero overlap with the spike in {\em both} variables.  We found---analytically, in numerical simulations, and in analysis of spectrogram correlations in Bengalese finch songs---that such successful detection is {\em always easier} from the joint- or the cross-covariance matrices than from  the individual self covariances. Thus, statistical methods exploiting cross covariances (PCA of the joint variable $Z$) or joint covariances (PLS between $X$ and $Y$), which we collectively call {\em simultaneous dimensionality reduction, SDR}, \cite{martini2024data} are  more data efficient than {\em individual dimensionality reduction, IDR}, which  start with self covariances (PCA of the individual variables, and then regressing $X$ and $Y$ principal components on each other). This resonates with the recent findings that SDR is more data efficient than IDR, analytically and numerically, in a variety of other linear and nonlinear methods \cite{zbontar2021barlow,radford2021learning,assran2023self,martini2024data,abdelaleem2023deep,abdelaleem2024simultaneous,abdelaleem2025accurate}. Recent work, developed simultaneously with and independently of ours, extends these results to detecting correlated signals in more than two variables using the joint method (and proposes an improvement to it) \cite{IEEEma, baharav2025stackedsvdsvdstacked}. Parenthetically, we note that we chose here not to explore methods that use both self- and joint-covariance matrices, such as CCA, since, for example, in its most straightforward form, CCA requires $\qx<1$ and $\qy<1$; the asymptotic performance of the method is then known \cite{Bouchaud2007}. In contrast, we are interested in the undersampled regime as more relevant to modern data science.)

While joint and cross covariances detect weaker signals than self covariances, neither is always superior to the other, and both have strengths and weaknesses. The joint covariance can detect an outlier even if the spike is extremely small in one of the two variables. This is not the case for the cross covariance, for which the product of the spike strengths, $ab$, must exceed the critical threshold. Yet, when the signal strengths of individual variables are similar, but dimensionalities are widely different, cross covariance bests the joint covariance. We confirmed numerically that this surprising result holds true for the latent feature model, which is a better model of actual data. 

At the very least, this suggests that different types of linear statistical methods, such as PLS or PCA on concatenated variables, should be used for data with different dimensionalities and different expected signal strengths, paralleling the investigation started in \cite{abdelaleem2024simultaneous}. This conclusion was also reached by another recent~\cite{léger2025highdimensionalpartialsquaresspectral} study using resolvent methods where it showed PLS-SVD could outperform individual PCAs. Overall, it is clear that principal component regression should never be used if the goal is to find correlations between two high-dimensional datasets with $O(1)$ linear latent variables mediating these correlations. Further, 
since the cross covariance approach becomes superior for dimensionally mismatched variables, where ``throwing out'' the poorly-sampled self covariance improves statistical power, it seems likely that there should be an intermediate linear method with an even better  performance, which would still rely on the self covariance of the better sampled variable, while ignoring the one of the undersampled one. 

It is also interesting to explore if all of these traditional and nontraditional methods are just special cases of a single Bayes-optimal approach \cite{keup2024optimal}, where the Bayes-optimal performance limits for multi-modal learning can be established using Approximate Message Passing (AMP). That analysis demonstrates that canonical spectral methods like PLS and CCA are sub-optimal, failing to reach the information-theoretic recovery thresholds that are achievable by more complex approaches. Another study using subgraph counting algorithm~\cite{li2025algorithmicphasetransitioncorrelated} identified that though the PLS threshold is strictly sub-optimal, it can still detect signals where individual PCA on $X$ and $Y$ may fail. Finally, one can also consider sequential approaches that have recently appeared in more complex multi-modal models involving mixed matrix–tensor observations~\cite{tabanelli2025computationalthresholdsmultimodallearning}, which connect naturally with curriculum inference strategies. Crucially, all of these approaches  involve leveraging the signal in one of the modalities to learn the signal in other one, and hence they still fall into the ``better together'' framework, emphasizing our main message that joint feature inference  should always be prioritized over simpler unimodal methods. 

Whether the intuition developed here translates to practical machine learning and statistical methods in a nonlinear context is an open question.   Self correlations based analysis---IDR---then corresponds to individual compression of $X$ and $Y$, presumably via nonlinear neural networks, and then seeking statistical relations between the compressed variables, again via optimizing some neural network. We already know that this approach is less data efficient than its SDR equivalents, namely compressing the two variables simultaneously, while  retaining as much information as possible between the compressed representations \cite{abdelaleem2023deep}. An analog of the joint covariance based method would then be using a concatenated critic to maximize the statistical dependencies between the compressed variables; the cross covariance methods would correspond to a separable critic (see \cite{abdelaleem2025accurate} and references therein). Whether a separable or a concatenated critic is better at detecting statistical dependencies between two datasets is still debated \cite{abdelaleem2025accurate}, and one can hope that the debate can be resolved similarly to our observation here:  mismatch of dimensionalities leads to a gradually increasing  advantage of a separable critic over a concatenated one. 

We hope that the analysis direction we open here, and especially the forthcoming investigations by the community of when joint or cross methods should be used for detecting correlations in paired signals, will be translated into new strategies for design of  detectors and the subsequent data analysis and compression for modern high-dimensional physics experiments, from large astronomical sensor arrays to optical imaging in biophysics.

\appendix
\section{Calculation of sub-components of the joint covariance} \label{app:projection}

As discussed in the main text, we want to evaluate $\xjol$ and $\yjol$, given our RMT calculation of $\zjol$. To do so, first recall how we have defined $\vxjoint$ and $\vyjoint$: first, we project $\vzjoint$ into the $X$ or $Y$ subspace, and then we normalize it. If we consider $\vx$ and $\vy$ to live in the full $\nx+\ny$ dimensional $Z$ space, we thus have
\begin{equation}
\xjol^2 = \frac{ \left|\vzjoint \cdot \vx \right|^2}{\sum_{i=1}^{N_X} \left|\vzjoint \cdot \hat{x}_i \right|^2,}
\end{equation}
with $\hat{x}_i$ a basis for the $X$ subspace (an equivalent formula holds for $Y$).

We first compute the overlap of $\vzjoint$ with an arbitrary unit vector $\hat{w}$. Any such vector can be written as
\begin{equation}
\hat{w} = (\hat{w} \cdot \vz) \vz + \sqrt{1-\left|\hat{w} \cdot \vz\right|^2} \hat{\delta},
\end{equation}
for some unit vector $\hat{\delta}$. We thus have  
\begin{align}
\left|\vzjoint \cdot \hat{w}\right|^2 &= \left| \hat{w} \cdot \vz \right|^2 \left|\vzjoint \cdot \vz \right|^2 \nonumber \\ &\quad+ \mleft(1-\left|\hat{w} \cdot \vz\right|^2\mright)  \left|\vzjoint \cdot \hat{\delta}\right|^2.
\end{align}

We can invoke rotational symmetry in the $\nx+\ny-1$ directions orthogonal to $\vz$ to find
\begin{multline}
\left\langle \left|\vzjoint \cdot \hat{w}\right|^2\right\rangle = \left| \hat{w} \cdot \vz \right|^2 \left\langle \zjol^2 \right\rangle \\ \quad+ \mleft(1-\left|\hat{w} \cdot \vz\right|^2\mright)  \left\langle\left|\vzjoint \cdot \hat{\delta}\right|^2\right\rangle \\
= \left| \hat{w} \cdot \vz \right|^2 \left\langle \zjol^2 \right\rangle + \frac{1-\left|\hat{w} \cdot \vz\right|^2}{\nx+\ny-1}.
\end{multline}
For the numerator, the first term is $O(1)$ and the second term is $O(1/N)$. Furthermore, the first term converges to its mean by standard RMT arguments, so the numerator converges to its mean.  We thus obtain
\begin{equation}
 \left|\vzjoint \cdot \vx\right|^2 =  \left|\vx\cdot \vz\right|^2 \zjol^2 = \frac{a^2}{a^2+b^2} \zjol^2.
\end{equation}
The denominator is trickier. Choose a basis in which $\vx = \hat{x}_1$. Then
\begin{multline}
\left\langle\sum_{i=1}^{\nx} \left|\vzjoint \cdot \hat{x}_i \right|^2 \right\rangle\\  
=  \frac{a^2}{a^2+b^2}\zjol^2 + \sum_{i=2}^{N_X} \frac{1-\zjol^2}{\nx + \ny - 1} \\
\approx  \frac{a^2}{a^2+b^2}\zjol^2 + \mleft(1-\zjol^2\mright) \frac{\nx}{\nx + \ny}.
\end{multline}
Again the first term converges to its mean by standard RMT arguments, while the second term is proportional to the projection of a vector onto a random extensive subspace, which has variance that goes to zero as $N\to\infty$, and thus also converges to its mean. Thus, the denominator converges to its mean, and
\begin{multline} \label{eq:split_overlap_X}
\xjol^2 \\\approx \frac{ \frac{a^2}{a^2+b^2}\zjol^2 }{ \frac{a^2}{a^2+b^2}\zjol^2 + \mleft(1-\zjol^2\mright) \frac{\nx}{\nx + \ny}}.
\end{multline}
For $Y$, we similarly have 
\begin{multline} \label{eq:split_overlap_Y}
\yjol^2 \\\approx \frac{ \frac{b^2}{a^2+b^2}\zjol^2 }{ \frac{b^2}{a^2+b^2}\zjol^2 + \mleft(1-\zjol^2\mright) \frac{\ny}{\nx + \ny}}.
\end{multline}

Note that the spike only entered into this calculation by determining the symmetry axis of the model. Thus, these results apply equally well to the latent feature model and the additive spike model.

\section{Joint overlaps in the latent feature model}\label{app:joint_latent}

The sample covariance matrix of the latent feature model is given by the multiplicative spike model in Eq.~(\ref{Eq.multiplicativejointspike}). The results for detecting the outliers and  the overlap of the eigenvector associated with the outlier eigenvalue and the spike are the same as those from the original BBP paper~\cite{BBP,wishart}. An outlier can be detected in joint covariance in the limit of very large matrix sizes if
\begin{equation} 
    c^2={a^2}+{b^2}\geq \ccrit^2=1+\sqrt{\qx+\qy}.\label{eq:joint_outlier_latent}
\end{equation}

For  $c^2\geq \ccrit^2$, the overlap is then 
\begin{equation}\label{eq:joint_overlap_latent}
     \zjol=        \sqrt{\left(1-\frac{\qx+\qy}{(c-1)^2}\right)/\left(1+\frac{\qx+\qy}{c-1}\right)},
\end{equation}
and zero otherwise.

As explained above, our results for converting the joint $Z$ overlap to the joint $X$ and joint $Y$ overlaps also apply for this model. Thus, the $X$ and $Y$ components of the joint overlap are obtained by substituting Eq.~(\ref{eq:joint_overlap_latent}) into Eq.~(\ref{eq:split_overlap_X}, \ref{eq:split_overlap_Y}).

\begin{acknowledgments} 
We thank Pierre Mergny and Lenka Zdeborova for extensive discussions and for sharing their results for the latent feature model. We thank Eslam Abdelaleem and K.~Michael Martini for stimulating discussions. This work was supported, in part, by the Simons Investigator award and NITMB grant to IN.
\end{acknowledgments}

\bibliography{references}

\end{document}